# Learner-Centered Analysis in Educational Metaverse Environments: Exploring Value Exchange Systems through Natural Interaction and Text Mining

**Yun-Cheng Tsai**
Department of Technology Application and Human Resource Development
National Taiwan Normal University
Taipei, Taiwan (R.O.C.)
pecu@ntnu.edu.tw

**Abstract**

This paper explores the potential developments of self-directed learning in the metaverse in response to Education 4.0 and the Fourth Industrial Revolution. It highlights the importance of education keeping up with technological advancements and adopting learner-centered approaches. Additionally, it focuses on exploring value exchange systems through natural interaction, text mining, and analysis. The metaverse concept extends beyond extended reality (XR) technologies, encompassing digital avatars and shared ecological value. The role of educators in exploring new technologies and leveraging text-mining techniques to enhance learning efficiency is emphasized. The metaverse is presented as a platform for value exchange, necessitating meaningful and valuable content to attract users. Integrating virtual and real-world experiences within the metaverse offers practical applications and contributes to its essence. This paper sheds light on the metaverse's potential to create a learner-centered educational environment and adapt to the evolving landscape of Education 4.0. Its findings, supported by text mining analysis, contribute to understanding the metaverse's role in shaping education in the Fourth Industrial Revolution.

**Keywords** Education 4.0 · learner-centered · educational metaverse · blockchain · value exchange · text-mining

## 1 Introduction

The advancement of virtual reality (VR) and augmented reality (AR) technologies has created new opportunities for educational experiences [1]. One emerging concept in this field is the educational metaverse, which surpasses traditional VR representations and offers seven-layered environments that enhance learner-centered experiences and facilitate value exchange systems [2].

However, the metaverse goes beyond the simple transformation of the physical world into a digital realm [3]. Virtual reality (VR) and augmented reality (AR) technologies are tools rather than being synonymous with the metaverse. It embraces the opportunity to meet new needs and establish conventions distinct to the digital world. Looking back at human history, tools are initially perceived as natural and necessary creations. It is essential to recognize that tools have played a significant role in shaping and creating humans. Human intelligence has evolved through continuous trial and error in tool usage and the accumulation of their value. Dissatisfaction with existing tools has driven changes in the tools themselves and the workforce. Even in the face of tool loss or insufficiency, humans persistently seek advancements and contemplate improvements. The evolution of these innovations has been instrumental in the development and progress of human civilization.

The metaverse concept extends beyond 3D or extended reality (XR), AR, VR, and mixed reality (MR) technologies. It encompasses digital avatars and the representation of shared ecological value. However, it is essential to acknowledge the potential existence of a recoverable paradox within the metaverse. This paradox arises from the belief that combining the variability of the virtual world with the physical world can enhance the governing value of the metaverse. For Instance, blockchain technology can revolutionize the credit system of universities, enabling the traceability of course values based on cryptocurrency.

This study conducts a learner-centered two-hour lecture course to challenge students' preconceived notions of an XR-dominated educational metaverse. Students actively understand and imagine the educational metaverse beyond XR concepts, emphasizing the value exchange system. Following the course, students are assigned to identify educational scenarios that cannot be realized within the metaverse. Their perceptions are collected through various materials, such as video recordings, written feedback, and audio files, for a comprehensive analysis.

Text mining techniques are increasingly used to analyze and assess learning outcomes in educational settings [4]. Through text mining, it becomes evident that students recognize the limitations of XR-restricted educational cases and develop a profound understanding of the role of blockchain in the educational metaverse. They also grasp the significance of the development and definition of the seven-layer theorem in shaping a learner-centered educational metaverse.

This paper explores the potential of analyzing educational metaverse environments from a learner-centered perspective through natural interaction and text-mining techniques. Based on the feedback and confirmation from students, we have become more convinced of the critical influence of "value exchange" in an education-centered approach. This has reaffirmed three beliefs for us:

1. Educational metaverse should explore the essence of value exchange rather than XR transformation alone. This has led to the misconception of prioritizing XR for the sake of XR, disregarding the actual purpose.
2. We should not negate the existence and necessity of the metaverse due to the disparities between virtual and physical realms. Instead, we should leverage value exchange to facilitate the transfer of value and amplify opportunities for value circulation through blockchain technology.
3. The concept of an educational utopia can only exist if we are entirely self-sufficient and do not rely on any form of exchange to fulfill basic needs. However, even the slightest dependency on others creates a need for value exchange. The educational metaverse aims to assist learners in rediscovering their value positioning, developing their worth, and being seen, utilized, and transferred. This process naturally leads them to identify the learning skills and knowledge they need to strengthen.

In conclusion, the educational metaverse seeks to empower learners to redefine their value and acquire the necessary skills and knowledge to enhance their worth. It recognizes the fundamental role of value exchange in the educational ecosystem, providing learners with opportunities to be identified, utilized, and have their value transferred. We can foster a more dynamic and effective educational environment by emphasizing value exchange.

This paper examines the potential advancements of self-directed learning within the metaverse, focusing on the implications for Education 4.0 and the Fourth Industrial Revolution. It emphasizes the significance of education, staying abreast of technological progress, and embracing learner-centered approaches. Furthermore, the study explores the value exchange systems in the metaverse by utilizing natural interaction, text mining, and analysis techniques. The metaverse is viewed as more than extended reality (XR) technologies, encompassing digital avatars and sharing ecological value. The role of educators is highlighted in their exploration of emerging technologies and their application of text-mining methods to enhance learning efficiency. The metaverse is presented as a platform that requires valuable and meaningful content to attract users. Integrating virtual and real-world experiences within the metaverse offers practical applications and contributes to its essence. This paper illuminates the metaverse's potential to foster a learner-centered educational environment and adapt to the evolving landscape of Education 4.0. The findings of this study, supported by text mining analysis, contribute to a deeper understanding of the metaverse's role in shaping education during the Fourth Industrial Revolution.

The following section examines the literature discussion on learner-centered analysis, educational metaverse based on a value exchange system, natural interaction, and text-mining. Section 3 presents the course design approach outlined in this paper, focusing on developing pedagogically meaningful activities. Section 4 discusses the outcomes and feedback from implementing the proposed method in the course. Subsequently, Section 5 outlines the limitations encountered during the study and provides suggestions for future research directions. Finally, our paper concludes in Section 6.

## 2 Literature

The literature reviewed in this section highlights the relationship between Education 4.0 and learner-centered approaches, emphasizing the integration of digital technologies and personalized learning experiences. Education 4.0 represents a paradigm shift in education, driven by rapid technological advancements, and aims to prepare learners for the Fourth Industrial Revolution. Learner-centered approaches prioritize individual needs, interests, and goals, promoting active engagement and self-directed learning. Integrating these two concepts is essential in designing effective educational environments.

Additionally, exploring learning outcomes through natural interaction and text-mining techniques offers new possibilities for assessing learners' progress and providing personalized feedback. Natural interaction allows learners to interact with educational content using intuitive interfaces, while text mining enables the analysis of written responses and discussions. Combining these techniques supports the development of adaptive learning systems and informs instructional design in a learner-centered Education 4.0 environment.

Furthermore, based on a value exchange system, the educational metaverse provides a platform for creating and exchanging valuable educational content and resources. This aligns with the broader concept of the metaverse as a platform for value creation and exchange. The educational metaverse offers interactive and immersive learning experiences, fostering more profound understanding, creativity, and learner collaboration.

Overall, the literature emphasizes the importance of learner-centered approaches, the integration of emerging technologies, and the value exchange system in shaping effective and engaging educational environments within the context of Education 4.0.

The following subsections provide detailed insights into the relationship between Education 4.0 and learner-centered approaches, the use of natural interaction and text mining in assessing learning outcomes, and the concept of the educational metaverse based on a value exchange system.

### 2.1 the Relationship between Education 4.0 and Learner-Centered

Education 4.0 represents a paradigm shift in education, driven by rapid technological advancements and the need to prepare learners for the Fourth Industrial Revolution. This transformation emphasizes the integration of digital technologies, personalized learning experiences, and learner-centered approaches [5]. The relationship between Education 4.0 and learner-centered approaches is a topic of significant interest and scholarly exploration [6].

Education 4.0 encompasses emerging technologies such as artificial intelligence, virtual reality, augmented reality, and big data analytics to enhance learning [7]. Learner-centered approaches, on the other hand, prioritize the individual needs, interests, and goals of learners, promoting active engagement and self-directed learning [8]. Integrating these two concepts is crucial to designing effective educational environments that empower learners to acquire relevant knowledge and skills in an increasingly digital and interconnected world.

Research has shown that learner-centered approaches within Education 4.0 can improve learning outcomes [9]. By placing learners at the center of the educational experience, educators can tailor instruction to meet their specific needs, interests, and learning styles. This approach promotes higher engagement, motivation, and critical thinking among learners, fostering a deeper understanding of the subject matter.

### 2.2 Exploring Learning Outcomes through Natural Interaction and Text Mining

Moreover, using natural interaction and text-mining techniques in assessing learning outcomes has gained attention in recent years [10]. Natural interaction refers to intuitive and immersive interfaces allowing learners to interact with educational content using gestures, speech, or other natural modalities. Text mining, on the other hand, involves extracting valuable information from large volumes of text data, enabling researchers to analyze and evaluate learners' written responses, discussions, and reflections [11].

The combination of natural interaction and text mining offers new possibilities for assessing learning outcomes in a learner-centered Education 4.0 environment. By analyzing learners' interactions, responses, and textual data, educators can gain insights into their progress, strengths, and areas for improvement. This information can inform instructional design, provide personalized feedback, and support the development of adaptive learning systems that cater to individual learners' needs [12, 13].

### 2.3 the Educational Metaverse Based on Value Exchange System

The educational metaverse involves creating virtual and natural interactive learning environments where learners can interact with digital content, engage in simulations, and collaborate with others in a shared space. The value exchange system within this metaverse entails creating and exchanging valuable educational content and resources among learners, educators, and other stakeholders [2, 14].

The concept of the educational metaverse based on the value exchange system aligns with the broader idea of the metaverse as a platform for value creation and exchange. It emphasizes the importance of meaningful and valuable content in attracting and engaging users within the educational context. Learners can have more interactive and immersive learning experiences, allowing for a deeper understanding of complex concepts and fostering creativity and collaboration.

## 3 Methodology

The methodology employed in this study involved implementing a pedagogical strategy known as "book discussion. [15]" After attending a lecture on a specific topic delivered by the speaker, students are divided into groups and assigned a related theme for further discussion. One month later, students are tasked with critically examining and identifying the impractical aspects of the theories presented by the speaker through their group presentations. These presentations served as a platform for students to challenge the speaker's assumptions and hypotheses by providing evidence and counterarguments. Following the student presentations, the speaker had the opportunity to clarify any misconceptions that may have arisen during the lecture by addressing the points raised in the students' reports. Additionally, the speaker could reassess their assumptions based on the students' rebuttals and refine any aspects of their original hypotheses subject to criticism.

### 3.1 Class Activity Design for Book Discussion

In this educational program, a total of 4 hours is allocated over one month. The first 2 hours are dedicated to the teacher providing a comprehensive introduction to the concept of the educational metaverse based on the value exchange system. During this time, students are encouraged to actively engage in the discussion, ask questions, and seek clarification.

1. Teacher to Students (2 hours): Following the initial 2-hour session, students are given 24 hours to reflect on the material covered and provide feedback. They are specifically prompted to break down complex information into constituent parts and examine their relationships. This exercise encourages critical thinking and a deeper understanding of the topic.
2. Teacher and Students are together (1 month): Once the feedback is collected, the teacher organizes the students into teams for further exploration. The teams collaborate and discuss various scenarios to identify aspects that cannot be executed within the educational metaverse. This group activity fosters teamwork, critical analysis, and problem-solving skills.
3. Student to Teacher (125 mins): Over the next month, the teams work together to delve into their assigned topics, considering different perspectives and potential limitations. They engage in thorough discussions and reflections, drawing on the knowledge gained during the initial lecture and incorporating their feedback. The aim is to present their findings and thoughts at the end of the month, showcasing their collective insights and conclusions. Following the last 2-hour session, students are given 24 hours to reflect on the material covered and provide their feedback. Each team has 25 minutes for presentations and discussions. The following is a breakdown of the time allocation for each team's activities:
   (a) A Team Presentation (20 minutes): Bring the students together for a team presentation. Please encourage students to think critically and formulate their thoughts and opinions about the teacher's themes, characters, or plot. Pose additional thought-provoking questions to stimulate deeper analysis and critical thinking.
   (b) Teacher's Reflection (5 minutes): The teacher summarizes the main points discussed after a team presentation. Teachers encourage students to reflect on their learning and insights from the teacher's main theory discussion.

There are a total of 24 master's students participating in this course. For the "Teacher and Students are together" task, the students are divided into four groups, and three groups comprise five members, while one group consists of four.

This structured approach gives students ample time to absorb the material, critically analyze it, engage in collaborative discussions, and present their reflections. By incorporating various stages of individual and group work, the program

Encourages active participation and comprehensive exploration of the educational metaverse based on the value exchange system.

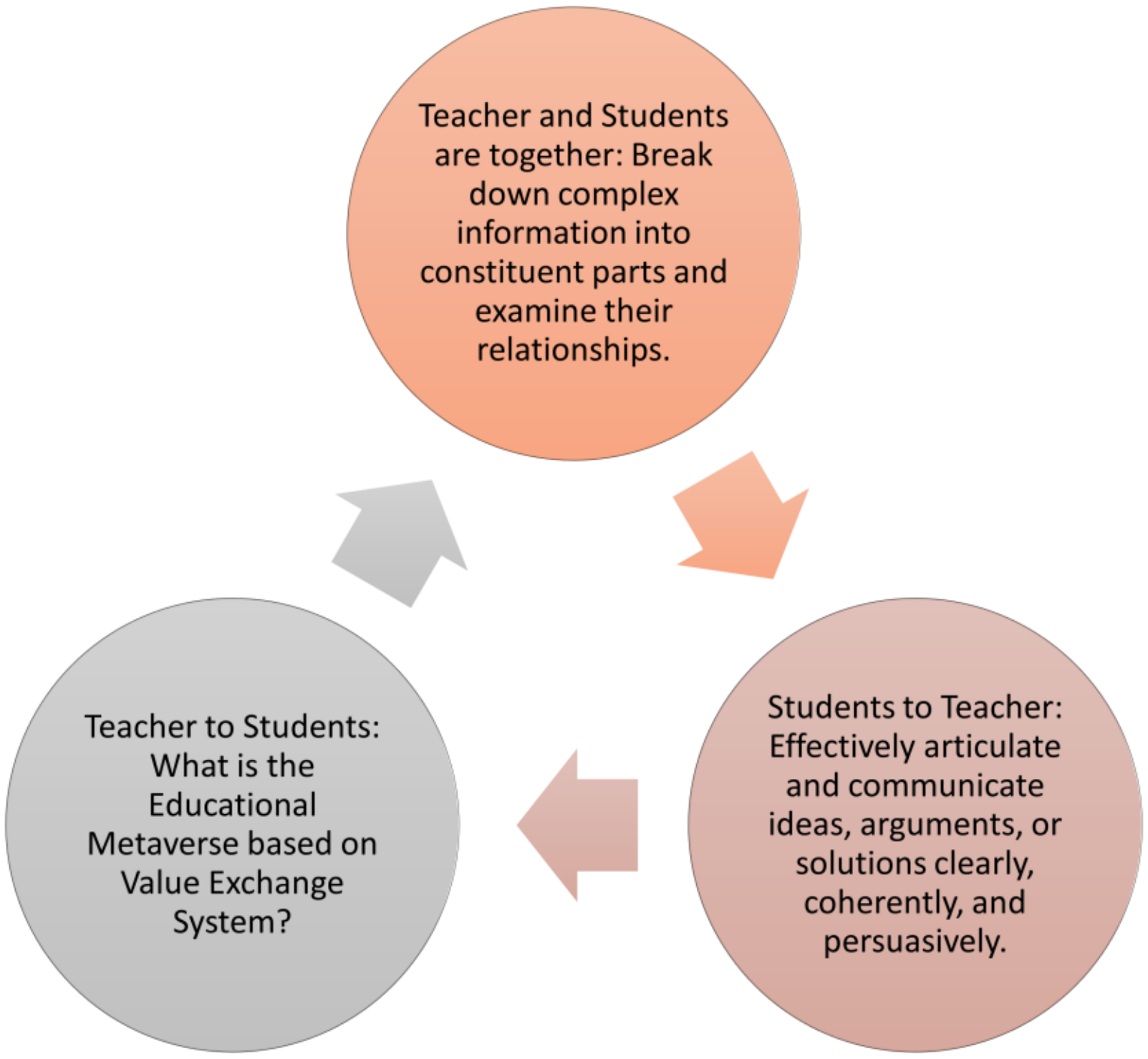

Figure 1: The cycle of speculation in the critical thinking process emphasizes the active engagement of students in evaluating the teacher's proposition regarding an educational metaverse focused on value exchange systems.

In Figure 1, the cycle of speculation in the critical thinking process emphasizes the active engagement of students in evaluating the teacher's proposition regarding an educational metaverse focused on value exchange systems. Notably, their efforts uncover the limitations present in the existing XR-based educational metaverse, as supported by the data they gather. This compelling evidence further strengthens the argument that the fundamental basis of an educational metaverse should revolve around the core concept of value exchange systems.

### 3.2 Extract Learning Outcomes from Verbatim Transcripts

Extracting learning outcomes from verbatim transcripts and using automated tools to transcribe audio recordings into written transcripts can provide a rich dataset for text-mining analysis. The following categories and content descriptions have been derived from the generated texts:

1. Teacher's Presentation: Transcriptions of the teacher's presentation on the educational metaverse and the value exchange system. Explanation of the educational metaverse concept, introduction to the value exchange system, and insights on the potential of blockchain in education.
2. Students' Feedback: Students provided textual feedback regarding the educational metaverse and the value exchange system. Positive comments, suggestions for improvement, reflections on the discussion, and opinions on the limitations of the XR-based educational metaverse.
3. Group Discussions and Students' Reflections: Students provided written reflections after a month, summarizing their thoughts and key takeaways from the course. Reflections on the importance of value exchange in education, insights gained from group discussions, and understanding of the seven-layer theorem in the educational metaverse. Ideas for enhancing the value exchange system, identifying educational scenarios that cannot be realized, and proposing solutions to overcome limitations.
4. Analysis and Insights: Textual analysis of the collected data to identify patterns, trends, and insights related to the educational metaverse and the value exchange system. Identification of common limitations in the XR-based educational metaverse, understanding the role of blockchain in enhancing the credit system, and exploration of the learner-centered aspects of the educational metaverse.

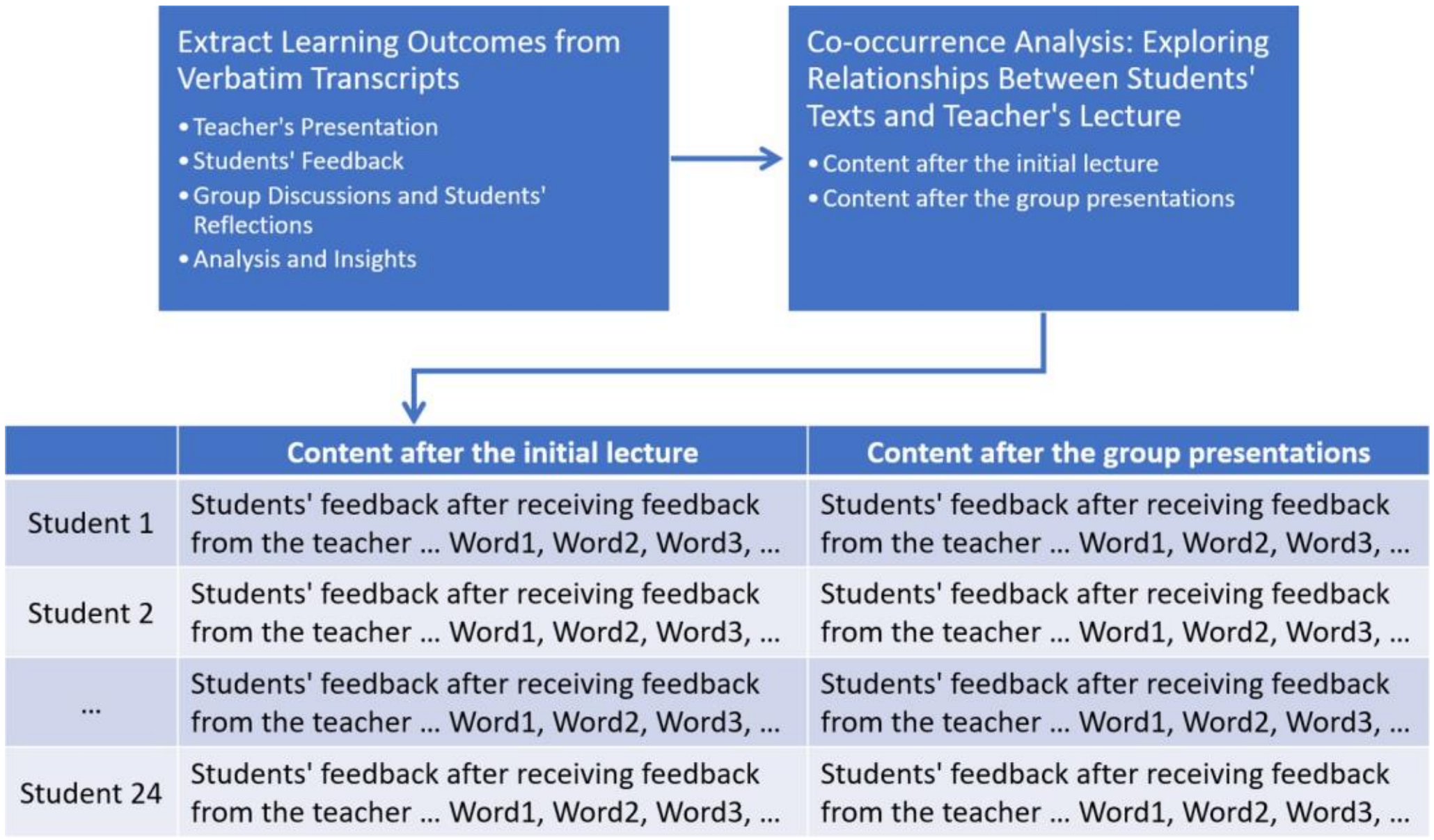

Figure 2: The Segmentation of Students' Texts and Teacher's Lectures.

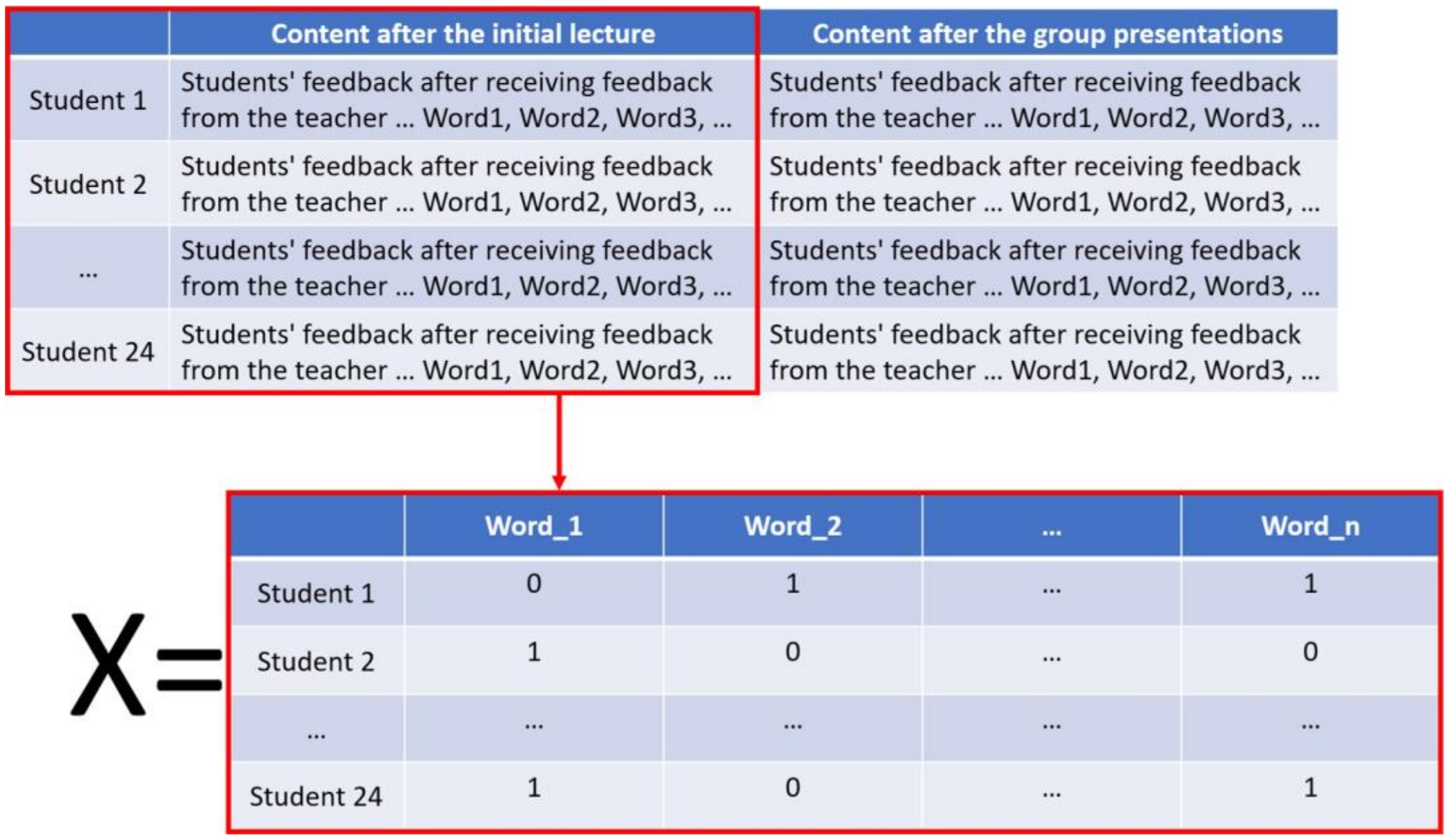

Figure 3: Create a term-to-document matrix by segmenting students' texts and the teacher's lecture.

### 3.3 Co-occurrence Analysis: Exploring Relationships Between Students' Texts and Teacher's Lecture

A text-mining approach has been designed to segment students' feedback into two categories to analyze these different types of texts and transcriptions. The first category includes the content provided immediately after the initial lecture, capturing their initial impressions and understanding. The second category comprises the content generated after the group presentations, including their reflections and responses to the feedback provided by the teacher. Figure 2 is the segmentation of the texts, and the following are the specific segments:

1. Content after the initial lecture: This segment includes students' initial responses and impressions of the teacher's lecture and their understanding and perspectives on the importance of value exchange in education.

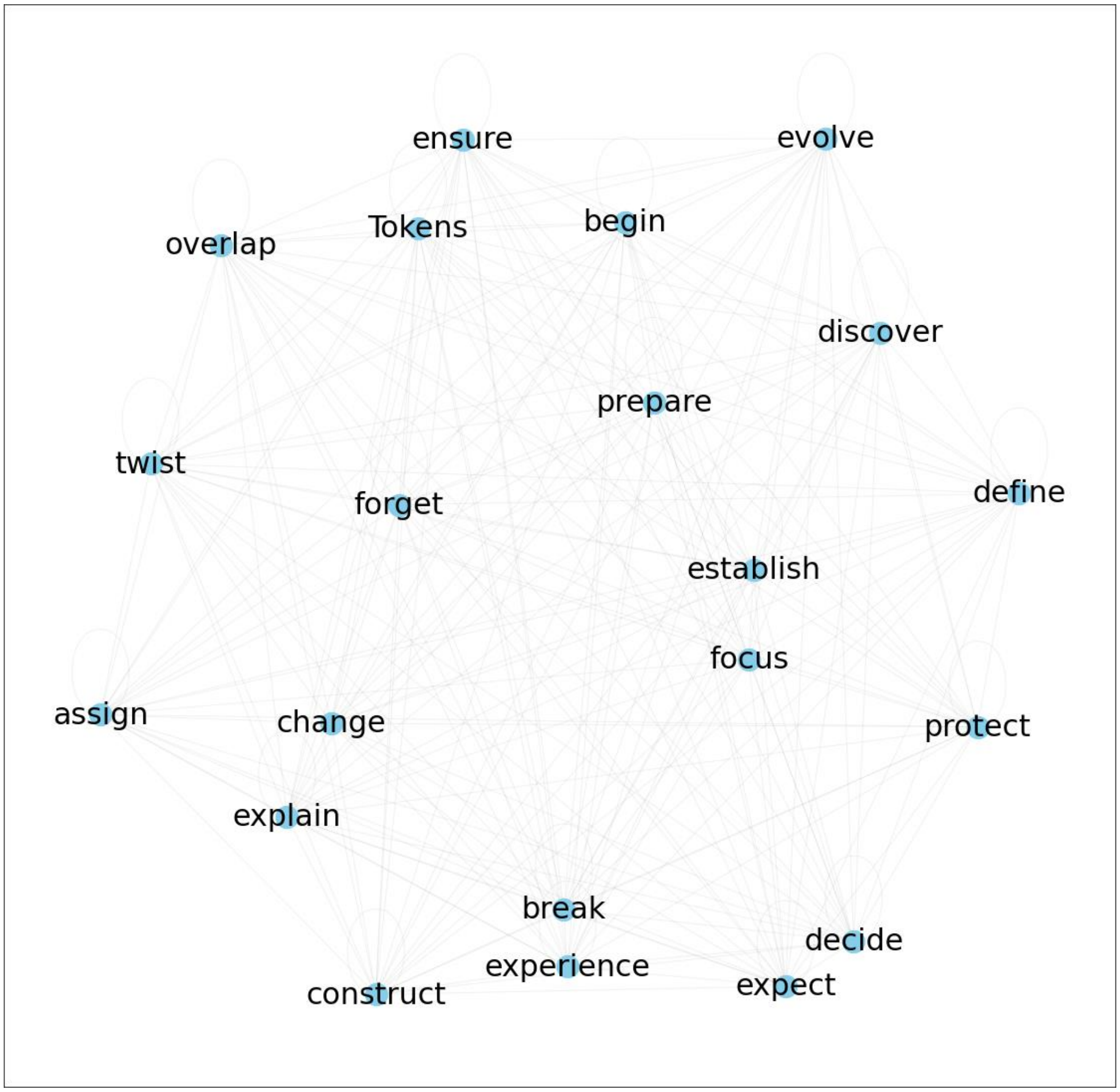

Figure 4: A co-occurrence graph from the Word Co-occurrence Matrix.

2. Content after the group presentations: This segment comprises students' observations and reflections following the group discussions and presentations. They can share new insights, perspectives, and experiences gained during the talks.

By categorizing the text into different categories and performing term extraction, you can create a term-to-document matrix, denoted as $X$ (see Figure 3), where each row represents a student, and each column represents a word extracted from the student's transcriptions. The entries in the matrix represent the frequency or presence of each expression in each student's text. To generate co-occurrence matrices [16], we can use two different approaches:

1. Word Co-occurrence Matrix ($X^\top X$): Transposing the term-to-document matrix $X$, we obtain $X^\top$, where each row represents a document, and each column represents a term. Multiplying $X^\top$ with $X$ yields the term co-occurrence matrix, where each entry represents the number of times two times co-occurs within the same document. This matrix captures the co-occurrence patterns of terms in one of the datasets.
2. Student Co-occurrence Matrix ($XX^\top$): Multiplying $X$ with its transpose $X^\top$ generates the student co-occurrence matrix, where each entry represents the number of terms two students share. This matrix captures the similarity or overlap in the understanding characteristics among the students.

When we have two categories, such as students' initial feedback and group presentations and feedback, and each type can generate two co-occurrence matrices, we will ultimately obtain four co-occurrence graphs for analysis. The four co-occurrence graphs represent the relationships and patterns of co-occurrence between words or phrases within each.

Category. By analyzing these graphs, we can gain insights into the connections and associations between terms or concepts within the students' feedback and their interactions with the teacher's lecture text.

For example, Figure 4 depicts a co-occurrence graph derived from a word co-occurrence matrix. This graph visualizes the relationships between words by representing them as nodes and using edges to indicate their co-occurrence patterns. Let's consider a corpus of news articles about technology. In the co-occurrence graph, each word corresponds to a node, and the connections between nodes represent the frequency or occurrence of these words appearing together in the text. Words that frequently occur together will have strong connections, while words that rarely co-occur will have weaker or no connections.

By examining the co-occurrence graph, we can observe which words frequently appear together in the text and identify clusters or groups of related terms. For instance, words like "technology," "innovation," and "digital" may form a closely connected cluster, indicating their thematic relevance. On the other hand, words like "technology" and "nature" may have weaker connections, suggesting less frequent co-occurrence and potentially indicating contrasting concepts.

Analyzing the structure of the co-occurrence graph can help researchers understand the semantic relationships between words and uncover underlying themes or topics within the text. This information can be helpful for task modeling, retrieval, or text summarization.

In summary, co-occurrence graphs visually represent word relationships in a text and offer insights into the clustering, association, or thematic relevance between words. By examining these graphs, researchers can gain a deeper understanding of the structure and content of textual data.

# 4 Results

The steps for conducting a co-occurrence analysis between students' reflections and the teacher's perspective are as follows:

1. Data preprocessing: Convert the students' reflection data into a suitable tokenization format, remove stopwords, and define keywords.
2. Feature extraction: Transform the preprocessed words into a document-to-term matrix, which serves as the basis for feature vectors representing students' views on specific terms.
3. Build a co-occurrence matrix:
   (a) Build a word co-occurrence matrix: Use the extracted feature vectors to construct a co-occurrence matrix. Each element represents the number of times two features appear together in the exact student text.
   (b) Build a student co-occurrence matrix: Utilize the extracted feature vectors to establish a co-occurrence matrix. Each element represents the number of times two members occur together using a specific term.
4. Construct co-occurrence networks: Based on the co-occurrence matrices from steps 3(a) and 3(b), create separate co-occurrence networks for word and student perspectives, where nodes represent features and edges define co-occurrence relationships.
5. Visualization and analysis: Visualize the co-occurrence networks and community structures using graph theory, enabling further analysis of patterns and trends in students' reflection changes.

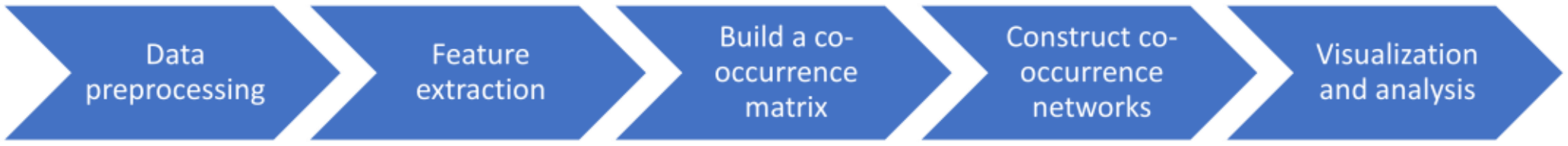

Figure 5: The steps for conducting a co-occurrence analysis between students' reflections and the teacher's perspective

Figure 5 illustrates the application of the analysis above steps, enabling a comprehensive understanding of the significant themes present in students' reflections during two instances, the interrelationships between these themes, and the changes observed over time. This analysis provides valuable insights into students' learning processes and their values and beliefs shifts.

Each co-occurrence graph visualizes the frequency and strength of co-occurrence between terms. Nodes in the graph represent individual terms, and the edges between nodes indicate co-occurrence relationships. The thickness or intensity of the edges reflects the strength of the co-occurrence. By examining the co-occurrence graphs, we can identify clusters or groups of related terms within each category. These clusters represent common themes, concepts, or ideas appearing.

| Students' IDs | Weight in Figure 7 | Students' IDs | Weight in Figure 9 |
|---|---|---|---|
| 19 | 506 | 8 | 1128 |
| 20 | 359 | 4 | 818 |
| 16 | 348 | 19 | 801 |
| 23 | 343 | 15 | 707 |
| 5 | 322 | 20 | 687 |
| 4 | 312 | 17 | 575 |
| 8 | 300 | 12 | 562 |
| 18 | 228 | 23 | 553 |
| 13 | 205 | 18 | 536 |
| 11 | 192 | 13 | 526 |
| 24 | 178 | 16 | 500 |
| 6 | 178 | 24 | 446 |
| 12 | 163 | 11 | 442 |
| 10 | 151 | 3 | 427 |
| 17 | 140 | 1 | 412 |
| 15 | 139 | 9 | 344 |
| 14 | 137 | 10 | 334 |
| 1 | 127 | 6 | 331 |
| 21 | 123 | 14 | 308 |
| 9 | 93 | 5 | 282 |
| 22 | 87 | 2 | 249 |
| 3 | 78 | 7 | 247 |
| 2 | 64 | 21 | 230 |
| 7 | 38 | 22 | 208 |

Table 1 shows the co-occurrence weights between students and teachers in the student co-occurrence matrix for two segmentation scenarios. The first two columns correspond to the consequences after the initial lecture, and the last two columns correspond to the weights after the group presentations.

In texts. Analyzing the structure and patterns of the graphs helps us understand the interrelationships and recurring topics within the students' feedback and their connection to the teacher's lecture. Analyzing these four co-occurrence graphs provides a total value of the links and associations in the students' feedback, highlighting the shared understanding, key concepts, and focus areas within each category. It assists in uncovering patterns, trends, and the overall coherence of the discussions and reflections related to the educational content.

Analyzing these co-occurrence matrices can provide insights into the co-occurrence patterns of words and the shared understanding characteristics among students. Researchers can apply techniques such as clustering, network analysis, or dimensionality reduction to explore the relationships and patterns within the matrices, thereby gaining a deeper understanding of students' learning outcomes and collective understanding.

Co-occurrence analysis is a powerful text technique that allows us to delve into the relationships between the texts of the 24 students and the teacher's lecture. By identifying co-occurring patterns, we can uncover associations and connections within the texts, shedding light on the student's level of understanding and engagement. In this analysis, we preprocess the texts by removing stopwords and punctuation, ensuring that the data is in a suitable format for examination. Subsequently, we construct a co-occurrence matrix, which captures the frequency of words appearing together in the students' texts andvalueteacher's lecture. The co-occurrence matrix serves as the foundation for our exploration. Examining the patterns within this matrix allows us to discern words or phrases that frequently co-occur in the texts. Such co-occurrences offer insights into the students' incorporation of ideas from the teacher's lecture into their reflections. If particular terms or concepts consistently appear together, it indicates the students' firm grasp and assimilation of those ideas. This analysis also enables us to identify key themes or topics that emerge across the texts, highlighting areas of shared understanding, agreement, or divergence between the students and the teacher. By comprehensively studying the co-occurrence patterns, we gain valuable insights into the level of engagement, comprehension, and assimilation of the lecture content among the students.

Figure 6 illustrates that the co-occurrence graph reveals three central nodes: "Us," "Metaverse," and "Value." The links originating from these central nodes demonstrate their dynamic interactions and relationships. Regarding "Us," we connect with the meta value by embracing Education 4.0, which harnesses cutting-edge educational technologies and digitalization to elevate the quality of learning and teaching experiences. Moreover, we engage with Value through the Education Metaverse, integrating values and ethical end-social responsibility principles into the learning journey. In this interaction, the Education Metaverse is seen as an ecosystem of value exchange, utilizing blockchain and digital tokens to verify and store learners' Value. This value exchange system allows learners to gain recognition and rewards through

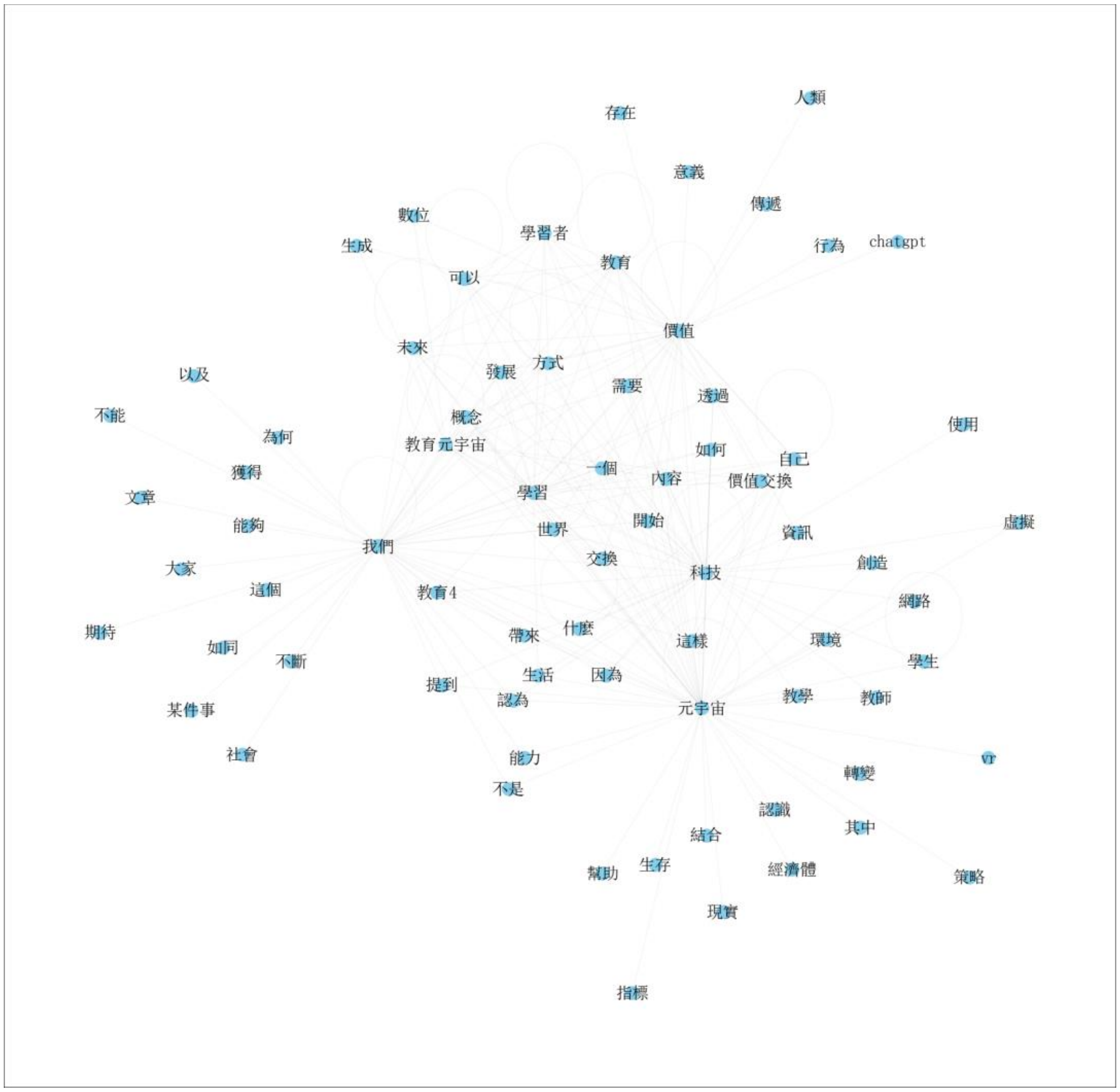

Figure 6: Word Co-occurrence Graph from the Content after the Initial Lecture.

Participation and contributions, actively engaging in the value circle. The interaction of the Education Metaverse has significant implications for the behaviors and meanings within the value circle. Through educational approaches and interactions among learners, they can explore and practice behaviors and definitions related to values, ethics, and social responsibility in the virtual world. The Education Metaverse provides a new learning environment where learners can experience and apply these value concepts and interact and collaborate with other learners. This article emphasizes the importance of Education Metaverse, Education 4.0, and value exchange. These concepts form an interactive system that promotes learners' value creation and growth. Additionally, the article reminds us to reflect and conduct further research on Education Metaverse while addressing potential challenges and issues.

Figure 8 illustrates a co-occurrence graph in which four central nodes are identified: "Metaverse," "Us," "Value," and "Experience." The Metaverse establishes a connection with Us through blockchain technology, facilitating various activities and interactions within the metaverse. Blockchain technology achieves this through data verification, storage, and transmission. The metaverse intersects with Value through creation, transfer, and environment. Within the metaverse, individuals can create and experience different values and cultures that can be shared and communicated. Value is linked to Experience through education and content. Education and content play crucial roles in the metaverse, providing learning opportunities and serving as carriers and conveyors of values. These interconnected elements form a value exchange system where Experience is critical. Through participation and experiences, we receive recognition and rewards for the Value we contribute while contributing our experiences and importance to the metaverse. Ultimately, this system completes a loop as experiences return to the blockchain, where our actions and deals are verified and stored. This loop allows us to continually participate, create, and exchange value within the metaverse. The co-occurrence

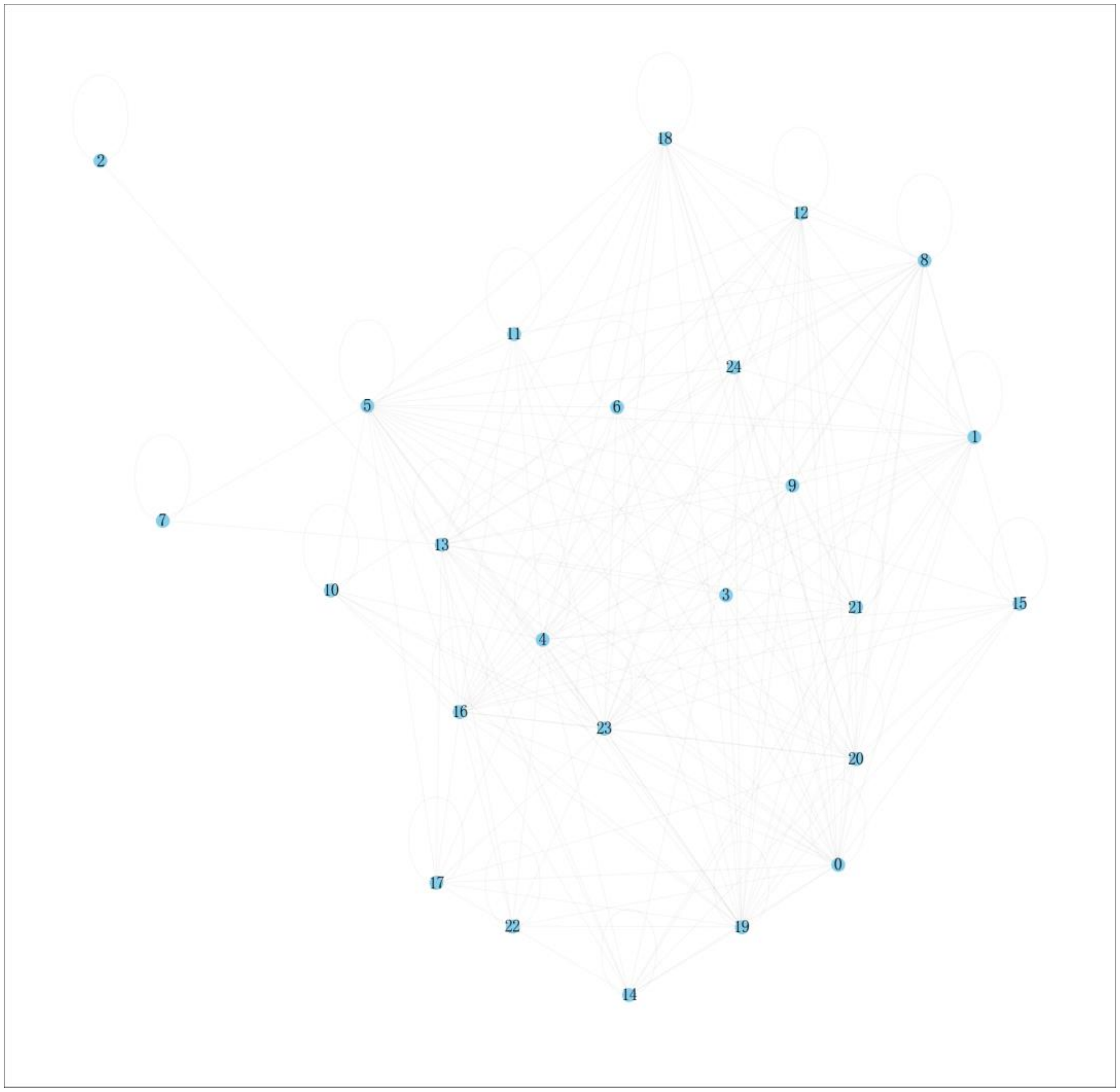

Figure 7: Student Co-occurrence Graph from the Content after the Initial Lecture.

The graph illustrates the relationships and interactions among the Metaverse, Us, Value, and Experience. This system facilitates value creation, sharing, and exchange while offering diverse learning and experiential opportunities.

The difference between Figure 6 and Figure 8 lies in their focus on the central nodes and their interconnections. The first paragraph describes a co-occurrence graph with central nodes of “Us,” “Metaverse,” and “Value,” emphasizing the role of Education 4.0 and the Education Metaverse. It highlights the value creation and growth students achieve through integrating these concepts. On the other hand, the second paragraph introduces "Experience" as an additional central node in the co-occurrence graph and emphasizes its relationship with Value. Based on these differences, it can be inferred that the students who wrote these texts have undergone some cognitive changes. They may have started to realize the crucial role of Experience in the Education Metaverse and its interconnectedness with Value. This shift in awareness might lead them to value practical experiences in the learning process and consider them integral to value creation and exchange. Additionally, they might contemplate how values can be conveyed and shared through education and content, recognizing the potential for practicing and applying these values in virtual environments. In writing these texts, these students have undergone a cognitive change in their understanding of the Education Metaverse, Education 4.0, and value exchange. They have become aware of the importance of Experience in value creation and exchange and have started contemplating how to embody these values within virtual learning environments. These cognitive changes reflect their deeper reflection and understanding of education and learning.

Figure 7 shows that 0 is the teacher and 1–24 are the students’ IDs. The student co-occurrence graph is from the content after the initial lecture. In Table 1, although the viewpoints of students 2 and 7 have lower co-occurrence with the teacher, careful examination of their descriptions reveals exciting insights. Student 2 emphasizes the curiosity sparked

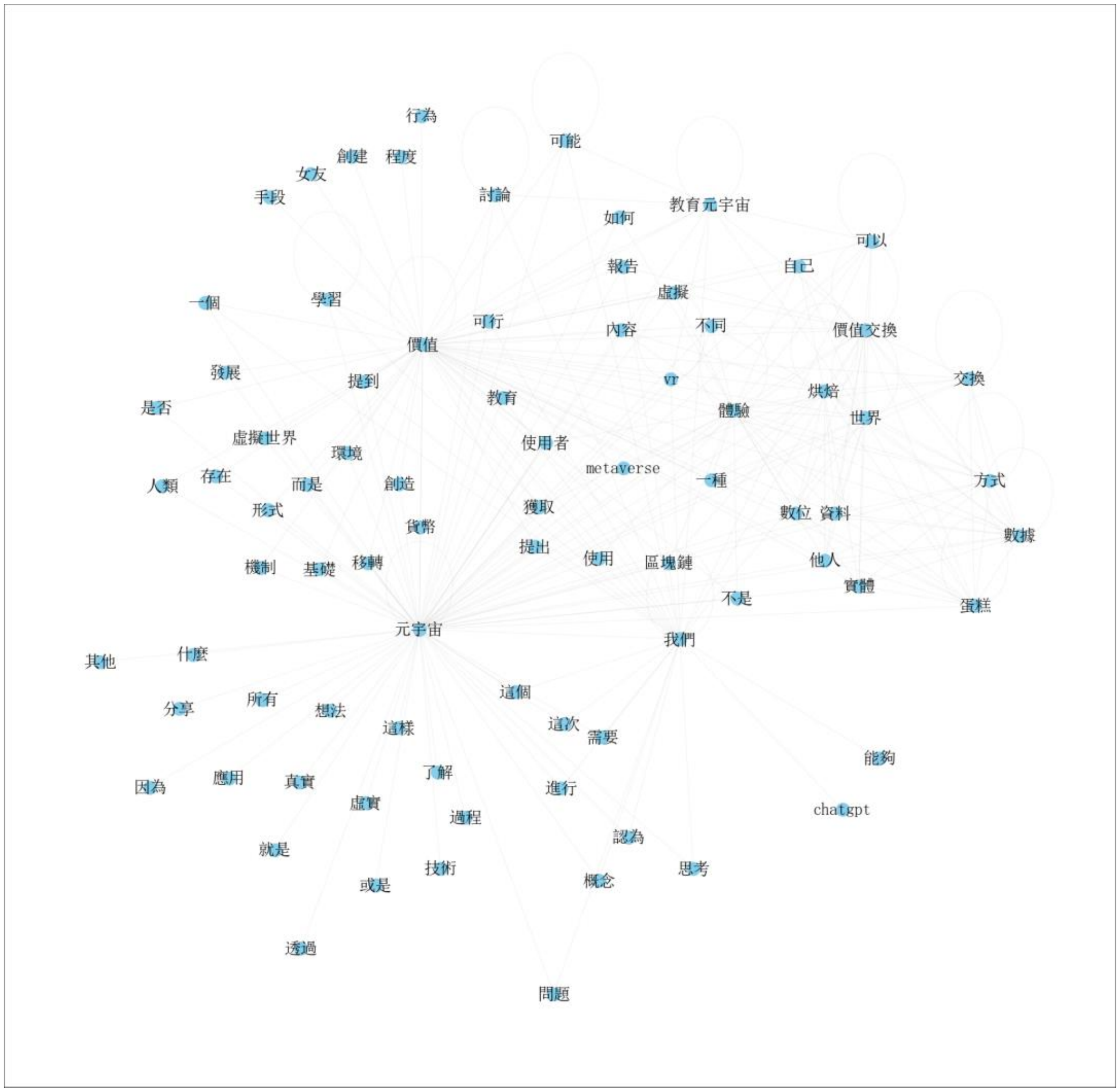

Figure 8: Word Co-occurrence Graph from the Content after the Group Presentations.

By understanding the teacher's life story, the pursuit of being valued, and self-directed and collaborative learning trends. On the other hand, student 7, in their conversation with the teacher, expresses a contradiction regarding the different statuses of education in the past and present society. They believe that learning should be connected to life experiences and that students should be the drivers of their knowledge. The emergence of the educational metaverse empowers teachers to serve as facilitators, enabling equal communication and fostering a deep understanding of the meaning and Value of learning for students. The article also highlights the importance of student-centered teaching approaches, leveraging technology and innovative methods. Educators should guide students to develop the necessary skills to face future challenges. The above observations highlight the nuanced perspectives of students, further emphasizing the relevance of the value exchange system in the educational metaverse.

Figure 9 shows that 0 is the teacher and 1–24 are the students' IDs. The student co-occurrence graph is from the content after the group presentations. Table 1 shows that the co-occurrence between students' and teachers' perspectives has significantly strengthened in the second round of feedback. All students' views now demonstrate a high degree of alignment with the teacher's viewpoint. This indicates that students clearly understand and can articulate the discourse surrounding an educational metaverse centered around a value exchange system. They can depict a narrative that closely aligns with the teacher's perspective. This suggests that students have embraced the concept of an educational metaverse where the value exchange system plays a central role. They have internalized its principles and can articulate their thoughts in a manner consistent with the teacher's discourse.

Co-occurrence weights are numerical values that quantify the strength or frequency of co-occurrence between entities or elements. In the context of the provided data, the co-occurrence weights represent the degree of association or

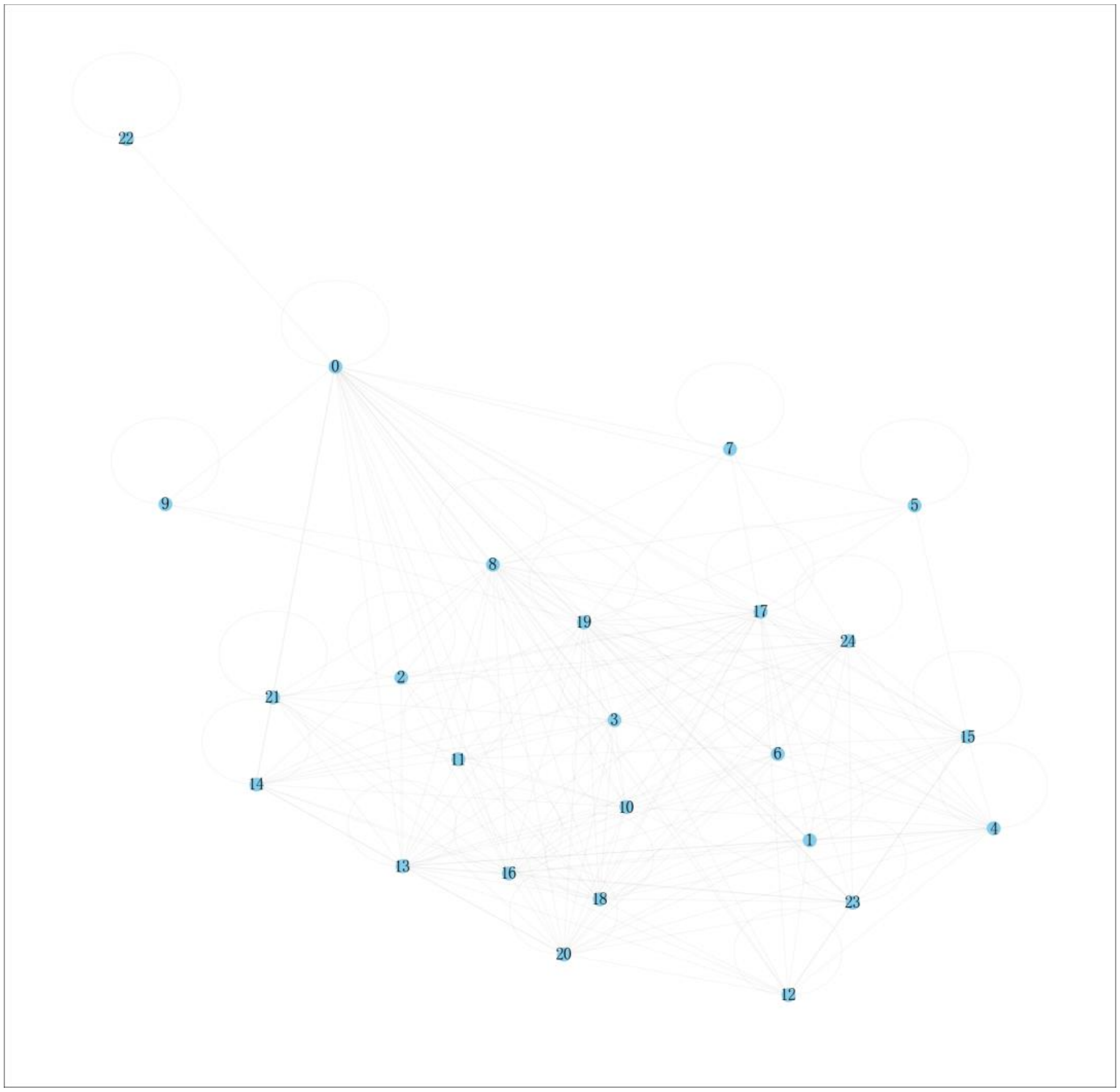

Figure 9: Student Co-occurrence Graph from the Content after the Group Presentations.

Alignment between the students' IDs and the teacher's viewpoint. Each entry in the data specifies the co-occurrence weight between a student's ID and the teacher's view. A higher weight indicates a stronger co-occurrence or alignment, suggesting that the student's viewpoint closely corresponds to the teacher's perspective.

These weights quantitatively measure how students' viewpoints agree with the teacher's. By analyzing and comparing these weights, we can identify the level of convergence or divergence between different individuals' perspectives within the given context. The co-occurrence weights can be used to identify patterns, similarities, or disparities in the responses and viewpoints of the students concerning the teacher's perspective. They provide valuable insights into the degree of agreement or alignment among the participants and help analyze the overall coherence or consensus within the group.

At the end of the month-long course, students were asked to provide written reflections, summarizing their thoughts and key takeaways. The considerations encompassed various aspects, including:

1. Importance of Value Exchange in Education: Students recognized the significance of value exchange systems in education. They reflected on how value exchange can foster meaningful and engaging learning experiences. Students identified the role of value exchange in promoting collaboration, creativity, and critical thinking.
2. Insights from Group Discussions: Students highlighted the value of group discussions and acknowledged the diverse perspectives and ideas shared during these discussions. Students appreciated the opportunity to learn from their peers and gain new insights into the educational metaverse.
3. Understanding of the Seven-Layer Theorem: Students demonstrated their understanding of the seven-layer theorem in the educational metaverse. They reflected on how the theorem shaped their perception of the Learner-centered environment. Students recognized the importance of the seven-layer theorem in creating a holistic and comprehensive educational metaverse.

4. Enhancing the Value Exchange System: Students brainstormed ideas for enhancing the value exchange system in the educational metaverse. They proposed strategies to ensure an equitable exchange of educational resources and content, and students explored innovative approaches to incentivize active participation and contribution within the value exchange system.
5. Identification of Unrealizable Educational Scenarios: Students critically analyzed the educational metaverse's limitations and constraints. They identified educational scenarios that cannot be realized within the current framework, and students reflected on the challenges and proposed potential solutions to overcome these limitations.

The group discussions and students' reflections provided valuable insights into their learning journey throughout the course. These reflections serve as a basis for further exploration and improvement in designing learner-centered educational environments and refining the value exchange system within the educational metaverse.

## 5 Limitations and Future Work

Despite the potential of using co-occurrence graphs and text mining techniques to analyze and understand the relationships between words in a given dataset, there are some limitations and areas for future work to consider.

1. Data quality and completeness: In the context of the metaverse, ensuring the accuracy and completeness of data used for analyzing learner interactions, virtual experiences, and value exchange systems is crucial. Future work can focus on developing methods to collect comprehensive and reliable data to construct accurate models and matrices for analysis.
2. Semantic understanding: While co-occurrence graphs provide insights into word relationships, the metaverse context requires a deeper understanding of words' semantic and contextual meaning. Future research can explore techniques to incorporate semantic analysis and natural language processing to capture the nuanced meaning and context-specific associations within the metaverse environment. Dynamic item analysis: The metaverse is a dynamic and evolving space. Future work can explore dynamic approaches to capture the changing nature of learner interactions, value exchange systems, and virtual experiences within the metaverse. This could involve real-time analysis of learner behavior and interaction patterns to adapt and personalize the learning experiences accordingly.
3. Integration of additional features: In addition to text mining techniques, future research can explore integrating elements such as sentiment analysis, sentiment mining, or user feedback analysis within the metaverse. This can provide a more comprehensive understanding of learners' experiences, emotions, and preferences, enabling the development of tailored interventions and personalized learning approaches.
4. Application in diverse educational contexts: While the current discussion focuses on self-directed learning in the metaverse, future work can expand the application of these concepts to other educational contexts within Education 4.0. This could include exploring how value exchange systems, learner-centered approaches, and technology integration can be implemented in various educational settings, such as online learning platforms, virtual classrooms, or collaborative learning environments.

By considering these aspects, researchers and practitioners can further extend the discussion and exploration of the potential developments of self-directed learning in the metaverse and its implications for learner-centered education in the context of Education 4.0 and the Fourth Industrial Revolution.

## 6 Conclusions

## Acknowledgments

We sincerely thank Professor Guan-Ze Liao from the Institute of Learning Sciences and Technologies at the National Tsing Hua University for his invaluable guidance and insightful discussions in his research methods and analysis seminar. The content of this course provided a crucial foundation for our study.